\documentclass[aps,superscriptaddress,groupedaddress,11pt]{revtex4}  
\usepackage{graphicx}  
\usepackage{dcolumn}   
\usepackage{bm}        
\usepackage{amssymb}   
\usepackage{amsmath}
\usepackage{float}
\usepackage{caption}
\usepackage{graphicx,psfrag,amsmath,subfigure}
\usepackage{color}
\usepackage{multirow}
\usepackage{sidecap}

\newcommand{\be}{\begin{equation}}
\newcommand{\ee}{\end{equation}}
\newcommand{\bea}{\begin{eqnarray}}
\newcommand{\eea}{\end{eqnarray}}
\newcommand{\bee}{\begin{eqnarray*}}
	\newcommand{\eee}{\end{eqnarray*}}

\begin{document}
	
	
	
	\hspace{5.2in} \mbox{CPHT-RR108.122019 }
	
\title{Diluting SUSY flavour problem on the Landscape}

\author{Emilian Dudas}
\affiliation{CPHT, Ecole Polytechnique, 91128 Palaiseau cedex, France}
\date{\today}

\author{Priyanka Lamba}
\affiliation{Centre for High Energy Physics, Indian Institute of
	Science, Bangalore- 560012, India}
\date{\today}

	\author{Sudhir K Vempati}
	\affiliation{Centre for High Energy Physics, Indian Institute of
		Science, Bangalore- 560012, India}
	\date{\today}	

		\begin{abstract}  
			We consider an explicit effective field theory example based on the Bousso-Polchinski 
			framework with a large number N of hidden sectors contributing  to  supersymmetry breaking. Each contribution comes from four form quantized fluxes, multiplied by random couplings.  
			The soft terms in the observable sector in this case become random variables, with mean values and standard deviations which are computable. We show that this setup naturally leads
			to a solution of the flavor problem in low-energy supersymmetry if N is sufficiently large. We investigate the consequences for flavor violating processes at
			low-energy and for dark matter. 
		\end{abstract}
		
		
		\maketitle
		\newpage
		\section{Introduction and Motivation} 
		\label{introsection}
%
%

Supersymmetry breaking in MSSM (Minimal Supersymmetric Standard Model) is introduced in terms of explicit
soft breaking terms. These are large in number $\sim$ 105, most of which violate flavor and CP symmetries. 
Phenomenologically there are strong constraints on the  flavor off-diagonal entries, requiring them to be suppressed
(compared to the flavour diagonal ones) by one  to several orders of magnitude. The flavor violation
constraints on the first two generations are significantly stronger compared to  the ones involving the third generation. 
These bounds are well documented in the literature \cite{Gabbiani:1996hi} (for reviews, see \cite{Dimopoulos:1995ju,Misiak:1997ei,Masiero:2005ua}).

To solve the problem with flavour violating soft terms, several solutions have been proposed. If supersymmetry
breaking is mediated purely  in terms of gauge interactions, the resulting soft terms would not contain any flavor
violation \cite{Dine:1993yw,Dine:1994vc, Dine:1995ag,Giudice:1998bp}. However, the discovery of the Higgs boson 
puts constraints on such models.  If the  Higgs boson is of supersymmetric origin, one would expect the mass of the lightest CP even  Higgs boson in Minimal Supersymmetric Standard Model (MSSM) to be rather
light.  This already puts severe constraints on the supersymmetric parameter space,  in particular that of the third generation up-type squarks (the stops): they are either required  to mix almost maximally or to be heavy,  between 3 to 4 TeV  (See for example, \cite{Bagnaschi:2018igf}). Several supersymmetry breaking models like minimal gauge mediation and its variations are  disfavoured in the light of the Higgs discovery
\cite{Draper:2011aa} or require a rather heavy spectra in the range of multi-TeV \cite{Bagnaschi:2018igf}.
		
In string or supergravity based models, it has been long known that it in general scenarios it is hard to escape 
flavour violation, unless some specific conditions are chosen \cite{Brignole:1993dj,Brignole:1997dp}.	 For example, 
if the K{\" a}hler potential of the matter fields is  canonical  and independent of the moduli/hidden
sector fields, one could expect an universal, flavor-independent form for the soft terms as in minimal Supergravity.   On the other hand, 
the problem can also be avoided if supersymmetry breaking is dominantly dilaton mediated \cite{Kaplunovsky:1993rd}. Other solutions include decoupling of the first two generations \cite{Cohen:1996vb}  or imposing flavour symmetries
(See for example, \cite{Antusch:2008jf,Pomarol:1995xc,Dine:1993np,Dudas:1995eq} and references there in).  
		
In the present letter, we would like to address these issues from a different point of view, insipred by the landscape of string theory vacua. We will consider a large number $N$ of sectors contributing to supersymmetry breaking. Large number of sequestered
 hidden sectors have also been considered recently in \cite{Cheung:2010mc,Cheung:2010qf,Benakli:2007zza}, in models with multiple (pseudo)goldstini.  Other works which have addressed supersymmetric soft spectrum phenomenology from the landscape following \cite{Denef:2004ze} include \cite{Baer:2019tee,Baer:2019xww,Baer:2017uvn,Baer:2017cck}.
 In particular a solution to flavour and CP problems in the landscape through heavy first two generations was
 proposed in \cite{Baer:2019zfl}.  
 
In the present work, we consider quantized four form fluxes {\ a} la Bousso-Polchinski \cite{Bousso:2000xa}.
		Each sector contributes in a quantized way, with a quantum that will be taken to be below the electroweak scale. Due to the large number of contributions, the observable soft terms become random variables 
		with Normal-type distributions around an average value. The setup has also the virtue of minimizing fine-tuning of the electroweak scale, due to the small contributions of each sector. 
		We find that our setup can address at the same time the flavor problem of low-energy SUSY by generating FCNC effects proportional to standard deviation of soft terms from their mean value, which are
		parametrically suppressed as $1/{\sqrt N}$. By performing a RG analysis from high to low-energy, the setup makes also concrete predictions for low-energy flavor observables. 
			
		The letter is organized as follows. In Section 2, we review the Bousso-Polchinski setup and follow it
	 in Section 3 with a review of four-forms fluxes. In the same section, we derive the soft terms and also show the
		impact on the flavour violating soft terms in the limit of large $N$. We also set up the boundary conditions for the
		scanning. In Section 4 we discuss the numerical results and present constraints from $K^0-\bar{K}^0$ oscillations and $\mu \to e +\gamma$. The framework also has interesting implications for dark matter which 
		we discuss briefly at the end of the section. We end with a small section of conclusions and outlook. 
		                                                                                                                                                 		
		\section{The Bousso-Polchinski setup}

The logic and setup we put forward is mainly originating from Bousso and Polchinski approach to the cosmological constant \cite{Bousso:2000xa}. We are regarding the implications of
		the string theory landscape for observable sector soft SUSY breaking terms. We will assume a large $N\gg1$ number 
	of SUSY-breaking sectors communicating through  gravitational couplings to the
		Supersymmetric Standard Model (SSM). 
Such models could naturally appear in string theory, where there may be several independent sources of supersymmetry breaking. 

Higher-dimensional operators and gravitational interactions lead to interactions between the moduli and the visible sector. One writes an effective action for the visible sector fields at a high scale, treating the hidden sector fields as non-dynamical background fields. This is justified if they are very heavy compared to the observable fields.  One then writes a set of renormalization group equations for the higher-dimensional operators and evolves them into the infrared, ignoring the hidden sector dynamics. Supersymmetry breaking F and D components of the background hidden sector fields then give rise to visible sector soft supersymmetry breaking parameters.\\
		Usually SUSY breaking is parametrized in terms of a single hidden sector field. This gives to a spectra at high scale (up to O(1) parameters) in terms of the scale of SUSY breaking F$_\alpha$. 	
		In our case, we consider instead MSSM interacting with N  hidden sectors at the
		Planck scale. In the  models we are considering, auxiliary fields of the hidden sector fields contain quantized four-form fluxes, with discrete charges contributing to supersymmetry
		breaking, as in the  Bousso-Polchinski solution to the cosmological constant problem.
		
		The main features of our framework are:
		\begin{itemize}
			\item Integer quanta parameterising the soft supersymmetry breaking contribution from each hid-
			den sector 
			\item Minimal number of parameters representing coupling between hidden sector fields and MSSM. Since these couplings depend on moduli vev's and interactions, we parametrize them by 
			random continuous parameters taking values inside a compact interval around zero.  
			\item Assume gravity mediation for simplicity.  A similar scan can be done for gauge mediation, although the details will be quantitatively different. 
			\item We consider a flat probability distribution of flux in each hidden sector. Since each flux is a random variable, due to the central limit theorem, this will lead to Normal-type distributions
			for the soft terms.
		\end{itemize}
		
		While we will impose the cancellation of the cosmological constant  \`a la Bousso-Polchinski, we do not necessarily use our framework 
		to address the cosmological constant problem. Instead we use the framework as a network of hidden 
		sectors each contributing individually to soft supersymmetry breaking.  		
\section{Four-forms  and fluxes}

Three form gauge potentials with (non-dynamical) four-form field strengths were considered longtime ago for addressing the cosmological constant problem \cite{Aurilia:1980xj,Witten:1983ux,Henneaux:1984ji,Brown:1987dd,Brown:1988kg,Duff:1989ah,Feng:2000if,dvali:2001sm}, the gauge hierarchy problem
\cite{Dvali:2003br,Dvali:2004tma,Herraez:2016dxn,Giudice:2019iwl}(see also \cite{Lee:2019efp}), the strong CP problem \cite{Dvali:2005an,Dvali:2005zk}, inflation \cite{Kaloper:2008fb,Kaloper:2011jz,Kaloper:2014zba,Dudas:2014pva} and supersymmetry breaking \cite{Farakos:2016hly}.  On the other hand,
it turned out to play an important role in the landscape of string theory compactifications \cite{Farakos:2017jme,Herraez:2018vae,Lanza:2019xxg} (for a recent review see e.g. \cite{Lanza:2019nfa}).  
Here we briefly review the main points of a theory containing three-forms with quantized form-forms field strengths. 

Let us start from a lagrangian containing some scalar fields $\varphi_i$ and three-form fields $C_{mnp}^\alpha$, with the action
\begin{equation}
{\cal S}_0 = \int d^4 x \ \{ - \frac{1}{2} (\partial \varphi_i)^2 - \Lambda_0 -  \frac{1}{2 \times 4 !} F_{mnpq}^{\alpha,2} +
\frac{1}{24} \ f_\alpha (\varphi_i) \ \epsilon^{mnpq} F_{mnpq}^\alpha \} \ , \label{shift1}
\end{equation}
where 
\begin{equation}
F_{mnpq}^\alpha = \partial_m C_{npq}^\alpha + {\rm 3 \ perms.} \ . \label{shift2}
\end{equation}
For future convenience we define
\begin{equation}
F^\alpha = \frac{1}{4 !} \epsilon^{mnpq} F_{mnpq}^\alpha \ , \ F_{mnpq}^\alpha = - \epsilon_{mnpq} F^\alpha  \ . \label{shift3}
\end{equation}
 
The lagrangian (\ref{shift1}) has actually to be supplemented with a boundary term 
\begin{equation}
{\cal S}_b = \frac{1}{6} \int d^4 x \ \partial_m \left( F^{mnpq}_\alpha C_{npq}^\alpha - f_\alpha (\varphi_i) \epsilon^{mnpq}
C_{npq}^\alpha \right)  \ . \label{shift4}
\end{equation}
The total action is
\begin{eqnarray}
&& {\cal S}  = {\cal S}_0 +{\cal S}_b =   \int d^4 x \ \{ - \frac{1}{2} (\partial \varphi_i)^2 - \Lambda_0 - \frac{1}{2 \times 4 !} F_{mnpq}^{\alpha,2} - \frac{1}{6} \ \epsilon^{mnpq} \partial_m f_\alpha (\varphi_i) \  
C_{npq}^\alpha \} \nonumber \\ 
&& + \frac{1}{6} \int d^4 x \ \partial_m \left( F^{mnpq}_\alpha C_{npq}^\alpha  \right)  \ .  \label{shift5}
\end{eqnarray}
A massless three-form gauge field in four spacetime dimensions has no on-shell degrees of freedom. As such, it
can be integrated out via its field eqs. 
\begin{equation}
\partial^m F_{mnpq}^\alpha \ = \  \ \epsilon_{mnpq} \partial^m f_\alpha (\varphi_i)  \ , \label{shift6}
\end{equation}
whose solution is given by
\begin{equation}
F_\alpha = -  f_\alpha (\varphi_i)  + c_\alpha \ , \label{shift7}
\end{equation}
where $c_\alpha$ is a constant, which is to be interpreted as a flux. It was argued in \cite{Bousso:2000xa} that $c_\alpha$
are quantized in units of the fundamental membrane coupling $c_\alpha = m_\alpha e$, fact that  has important consequences for the landscape of string theory.
After doing so, the final lagrangian takes the form
\begin{equation}
{\cal S} = \int d^4 x \ \{ - \frac{1}{2} (\partial \varphi_i)^2 - \Lambda_0 - \frac{1}{2} \sum_\alpha (f_\alpha (\varphi_i)-c_\alpha)^2 \} \ . \label{shift8}
\end{equation}
The final resulting cosmological constant is therefore scanned by the flux
\begin{equation}
\Lambda =  \Lambda_0 + \frac{1}{2} \sum_\alpha (f_\alpha (\varphi_i)-c_\alpha)^2  \ . \label{shift9}
\end{equation}
Notice that the boundary term ${\cal S}_b$ is crucial in obtaining the correct action. Ignoring it leads
to the wrong sign of the last term in  (\ref{shift8}), fact that created confusion in the past. 
\subsection{Supersymmetric formulation}

The embedding of four-form fluxes in supersymmetry and supergravity proceeds by introducing three-form multiplets, defined as the real superfields
\cite{Gates:1980ay,Burgess:1995kp,Burgess:1995aa,Binetruy:1995hq,Binetruy:1996xw,Binetruy:2000zx,Groh:2012tf} 
\begin{eqnarray}
&& U_\alpha = {\bar U}_\alpha = B_\alpha + i (\theta \chi_\alpha - {\bar \theta} {\bar \chi}_\alpha) + \theta^2 {\bar M}_\alpha +
 {\bar \theta}^2 { M_\alpha} + \frac{1}{3} \theta \sigma^m {\bar \theta} \epsilon_{mnpq} C^{npq}_\alpha
+ \nonumber \\ 
&& \theta^2 {\bar \theta}  (\sqrt{2} {\bar \lambda}_\alpha + \frac{1}{2} {\bar \sigma}^m \partial_m \chi_\alpha)
+ {\bar \theta}^2 \theta  (\sqrt{2} {\lambda}_\alpha - \frac{1}{2} {\sigma}^m \partial_m {\bar \chi}_\alpha)
+ \theta^2 {\bar \theta}^2 (D_\alpha-\frac{1}{4} \Box B_\alpha) \ . \label{susy1}
\end{eqnarray}
The difference between $U_\alpha$ and a regular vector superfield $V$ is the replacement of the vector potential
$V_m$ by a three-form $C^{npq}_\alpha$. 
In order to find correct kinetic terms, the analog of the chiral field strength superfield $W_{\alpha}$
for a vector multiplet is replaced by the chiral superfield \cite{Gates:1980ay}
\begin{equation}
T_\alpha = - \frac{1}{4} {\bar D}^2 U_\alpha \quad , \quad T_\alpha (y^m,\theta) = M_\alpha + \sqrt{2} \theta \lambda_\alpha +
\theta^2 (D_\alpha + i F_\alpha) \ , \label{susy2}
\end{equation}
with $F_\alpha$ defined as in (\ref{shift3}). The definition (\ref{susy2}) is invariant under the gauge transformation $U_\alpha \to U_\alpha - L_\alpha$, where $L_\alpha$ are linear multiplets. Correspondingly, lagrangians expressed as a function of $T_\alpha$ will have this gauge freedom.   One can therefore choose a gauge in which $B_\alpha=\chi_\alpha=0$  in  (\ref{susy1}) and the physical fields are 
complex scalars $M_\alpha$ and Weyl fermions  $\lambda_\alpha$. 

Notice that for the purpose of finding the correct on-shell lagrangian and scalar potential, there is a simpler formulation in which $T_\alpha$ are treated as  standard chiral superfields with 
$D_\alpha+iF_\alpha$ as 
auxiliary fields, no boundary terms are included, but the superpotential of the theory is changed
according to \cite{Binetruy:1996xw,Binetruy:2000zx,Groh:2012tf}
\begin{equation}
W (\phi_i, T_\alpha) \to W' (\phi_i, T_\alpha) \ = \ W (\phi_i, S_\alpha) + i c_\alpha T_\alpha \ , \label{susy07} 
\end{equation}     
where $c_\alpha$ are the quantized fluxes. The linear terms in the superpotential shift linearly the auxiliary fields. In supergravity, the (F-term) scalar potential can be written as ($M_P=1$ in what follows)
\begin{equation}
V= K_{\alpha \bar \beta} F^{\alpha}  F^{\bar \beta} - 3 m_{3/2}^2 \quad , \quad {\rm where} \quad  F^{\alpha} = e^{\frac{K}{2}} K^{\alpha \bar \beta} \overline{D_\beta W} \ .    \label{susy08}
\end{equation}
The linear flux terms shift therefore the auxiliary fields according to
\begin{equation}
F^{\alpha} = e^{\frac{K}{2}} K^{\alpha \bar \beta} \overline{D_\beta W'} = e^{\frac{K}{2}} K^{\alpha \bar \beta} \overline{D_\beta W} - i  e^{\frac{K}{2}} K^{\alpha \bar \gamma}
 ({\bar c}_{\bar \gamma} + K_{\bar \gamma} {\bar T}_{\bar \beta} {\bar c}_{\bar \beta} )  \ ,    \label{susy09}
\end{equation}
 leading to a scanning of the cosmological constant.
 \subsection{Soft terms in supergravity}
 
 We start from a supergravity lagrangian containing hidden sector (moduli) fields $T_\alpha$, whose auxiliary fields contain the four-form fluxes we introduced previously, coupled to matter fields
 called $Q_i$ in what follows. The Kahler potential and superpotential are defined by
 \begin{eqnarray}
 && K = {\hat K} (T_\alpha, {\bar T}_{\alpha}) + K_{i \bar j} (T_\alpha, {\bar T}_{\alpha}) Q^i {\bar Q}^{\bar j} + \frac{1}{2} \left( Z_{i j} (T_\alpha, {\bar T}_{\alpha}) Q^i {Q}^{ j} + {\rm h.c.} \right) 
 \ , \nonumber \\
 && W = {\hat W} (T_\alpha) + \frac{1}{2}  {\tilde \mu}_{i j} (T_\alpha) Q^i {Q}^{ j} +  \frac{1}{3}  {\tilde Y}_{i jk} (T_\alpha) Q^i {Q}^{ j} Q^k + \cdots    \ .   \label{soft1}
\end{eqnarray}
The low-energy softly broken supersymmetric lagrangian is defined by the superpotential and soft scalar potential
\begin{eqnarray}    
&& W_{\rm eff} = \frac{1}{2}  {\mu}_{i j} Q^i {Q}^{ j} +  \frac{1}{3}  {Y}_{i jk}  Q^i {Q}^{ j} Q^k \ , \nonumber \\ 
&&  {\cal L}_{\rm soft} = - m_{i \bar j}^2 q^i q^{\bar j} - \left( \frac{1}{2}  {B}_{i j} q^i {q}^{ j} +  \frac{1}{3}  {A}_{i jk}  q^i {q}^{ j} q^k + \frac{1}{2} M_a \lambda_a \lambda_a + {\rm h.c.} \right) \ ,   \     \label{soft2}
\end{eqnarray} 
where $Y_{ijk} = e^{K/2}  {\tilde Y}_{i jk}$. After imposing the cancellation of the cosmological constant,  the various soft terms and the supersymmetric masses are given by 
\cite{Kaplunovsky:1993rd,Brignole:1993dj,Brignole:1997dp,Ferrara:1994kg,Dudas:2005vv}
\begin{eqnarray}
&& M_a = \frac{1}{2} g_a^2 F^{\alpha} \partial_\alpha f_a \ , \nonumber \\
&& m_{i  \bar j}^2 = m_{3/2}^2 K_{i \bar j} - F^{\alpha}  F^{\bar \beta} R_{i \bar j \alpha \bar \beta} \quad , \quad {\rm where} \quad R_{i \bar j \alpha \bar \beta} = \partial_\alpha \partial_{\bar \beta}
K_{i \bar j}  - K^{m \bar n} \partial_\alpha K_{i \bar n} \partial_{\bar \beta} K_{m \bar j} \ , \nonumber \\
&& A_{ijk} = (m_{3/2} - F^{\alpha} \partial_\alpha \log m_{3/2} ) Y_{ijk} + F^{\alpha} \partial_\alpha Y_{ijk} - 3 F^{\alpha}  \Gamma_{\alpha (i}^l Y_{ljk)} \ ,    \label{soft2}
   \end{eqnarray} 
where we have introduced also the Kahler connexion
\begin{equation}
\Gamma_{IJ}^K = K^{K \bar L} \partial_I K_{J \bar L} \ .    \label{soft3}
\end{equation}  $B_{ij}$ terms are not displayed since they will not be scanned in what follows. Similarly the
$\mu$ term is determined by the radiative electroweak symmetry breaking conditions at the weak scale.

 \subsection{Scanning soft terms and gravitino mass}
 
 Taking into account the scanning of auxiliary fields from four-forms fluxes, in what follows we use the simplified scanning
 \begin{equation}
F_\alpha = m_\alpha {\tilde m} M_P \ , \label{scan1}
 \end{equation}
where $m_\alpha$ are integers and $\tilde m M_P$ is the quantum of scanning. Taking into account the cancellation of the cosmological constant, and setting the matter fields wavefunctions in a canonical form, the formulae for the scanning will be taken to be 
\begin{eqnarray}
&& m_{3/2} = \tilde m (g_0 + \sum_{\alpha}g_\alpha m_\alpha) \quad , \quad m_{3/2}^2= \frac{1}{3} \sum_{\alpha=1}^N \frac{F_\alpha^2}{M_P^2} = \frac{1}{3} {\tilde m}^2 \sum_{\alpha} {m_\alpha^2}  \ , \ \nonumber \\
&& (m_0^2)_{i \bar j} = m_{3/2}^2 \delta_{i \bar j}  +  {\tilde m}^2 \sum_{\alpha}  d_{\alpha, i \bar j}  m_\alpha^2 \ , \ \nonumber \\
&& M_{1/2}^a = {\tilde m} \sum_{\alpha}  s_{\alpha}^a  m_\alpha \ , \ \nonumber \\
 && A_{i jk} = m_{3/2} y_{i jk}  +  {\tilde m}  \sum_{\alpha}  a_{\alpha, i jk}  m_\alpha \ , \  \label{scan2}
 \end{eqnarray}
 where $\alpha = 1 \cdots N$, $-M \leq m_\alpha \leq M$ (integers), $-d_0 \leq a_\alpha, d_\alpha,s_\alpha,g_\alpha \leq d_0$ (continuous). These last couplings are taking to be continuous in order
 to take into account the couplings of the hidden sector fields with the MSSM ones, which are dependent on the hidden sector vev's and interactions. Note that the soft terms defined above are in the so-called super CKM basis which is important for the flavor discussion below. 
 
The scanning of the gravitino mass, combined with the cancellation of the cosmological constant (the Deser-Zumino relation) implies a constrained among the fluxes
\begin{equation}
g_0^2 + 2 g_0 \sum_\alpha g_\alpha m_\alpha + \sum_{\alpha,\beta}  g_\alpha g_\beta  m_\alpha  m_\beta  = \frac{1}{3} \sum_\alpha m_\alpha^2 \ . \label{scan3}
\end{equation}
Taking the average value of (\ref{scan3}) this implies in particular 
\begin{equation}
g_0^2 = \sum_\alpha \overline{(1/3- g_\alpha^2) m_\alpha^2}\simeq \frac{N}{9}(1-d_0^2)M^2 \quad , \quad \overline{m_{3/2}} = {\tilde m}  g_0 \sim O(\sqrt{N}) {\tilde m}  \ . \label{scan4}
\end{equation} where we have used the large flux limit  $M \gg1$. We can also compute
\begin{equation}
\overline{m_{3/2}^2} = \frac{\tilde m^2}{3} \sum_\alpha \overline{m_\alpha^2}  \sim    \frac{1}{9} N M^2 {\tilde m^2} \ , \label{scan4}
\end{equation}
The mean values of the soft terms are therefore computed to be
\begin{equation}  
\overline{(m_0^2)_{i \bar j}} = \overline{m_{3/2}^2} \delta_{i \bar j} ,\quad \quad \overline{A_{ijk}} = \overline{m_{3/2}} y_{i jk}, \quad  \quad   \overline{M_{1/2}^a} = 0  \ .  \label{scan5}
\end{equation}
There are two type of averages : one over the flux quanta $m_\alpha$ and the other over the (continuous) couplings $d_\alpha$. Being independent variables, one can use formulae of the type
\begin{equation}
\overline{f_1 (d_\alpha) f_2 (m_\alpha)} = \overline{f_1 (d_\alpha)} \times \overline{ f_2 (m_\alpha)}  = \frac{1}{2d_0} \int_{-d_0}^{d_0} dx f_1 (x)  \ \times \
\frac{1}{2M+1} \sum_{m_\alpha=-M}^M f_2 (m_\alpha)  \ . \label{scan6}
\end{equation}
By using such formulae, one finds
\begin{eqnarray}
&& (\Delta m_0^2)^2 = (\Delta m_{3/2}^2)^2 + {\tilde m}^4 \sum_\alpha \overline{d_\alpha^2 m_\alpha^4} \simeq \frac{NM^4}{15}\tilde{m}^4\left(\frac{4}{27} + d_0^2 \right) \ , \nonumber \\
&& {\rm where} \quad (\Delta m_{3/2}^2)^2 = \frac{{\tilde m}^4}{9} \sum_\alpha [\overline{m_\alpha^4} - (\overline{m_\alpha^2})^2 ]  \ . \label{scan7}
\end{eqnarray}
Consequently, one finds
\begin{equation}
(\delta_{ij})_{LL/RR} \equiv
\frac{\delta m_0^2}{\overline{m_0^2}} \simeq \frac{1}{\sqrt{N}} \sqrt{ \frac{1}{5}\left(4 + 27d_0^2\right)} \ . \label{scan8}
\end{equation}
The off-diagonal entries, which have zero average values, are governed by the standard deviation $\delta m_0^2$. One concludes then that they are suppressed compared to the
diagonal entries. For a large number of hidden sector $N \geq 10^{6} $, the flavor problem of MSSM is therefore solved. While this discussion is considering the flavor violating entries at the supergravity scale, in practice at the
weak scale, as we will see in the next section, $N\sim 100$ would be sufficient to absolve  strong constraints from  $\Delta m_K$. For the constraint from $\mu \to e +\gamma$, however, $N\sim 100$ is not sufficient 
and a larger value of $N$ should be chosen.  

It should be noted however that the above discussion is pertaining to definition of $\delta_{ij}$ at the high scale. 
At the weak scale, for the leptonic sector (in the absence of right handed neutrinos), there is no significant 
change in the mean values, where as for the hadronic (squark) sector, due to the large gluino contributions
to the squark masses in RG running, the $\delta_{ij}^{q,u,d}$ would be further suppressed by a factor  from $7$ up to 
an order of magnitude. 

For the gaugino masses, one finds
\begin{equation}
\Delta M_{1/2}^2 = {\tilde m}^2 \sum_\alpha \overline{s_\alpha^2 m_\alpha^2} \simeq N {\tilde m}^2 \frac{d_0^2 M^2}{9}  \ . \label{scan9}
\end{equation}
Therefore one finds the standard deviation
\begin{equation}
\Delta M_{1/2} = d_0 \sqrt{ \overline{m_{3/2}^2}}  \ . \label{scan10}
\end{equation}
For A-terms, let us consider for definiteness
\begin{equation}
A^u = m_{3/2} y_D^u + {\tilde m} \sum_{\alpha}a_\alpha^u m_{\alpha} \ , \label{scan11}
\end{equation}
where in the mass basis for fermions and  scalar-fermion-gaugino couplings are diagonal, $y_D^u $ is diagonal in the flavor space. If $a_\alpha^u \sim y^u \tilde{a}_\alpha $, then one expects the flavor violation 
in this case to be under control. However, if this is not the case, we can use the same arguments as above. 
One then finds the standard deviation
\begin{eqnarray}
&& (\Delta A^u)^2 = \overline{m_{3/2}^2} (y_D^u)^2 + {\tilde m}^2 \sum_\alpha  \overline{(a_\alpha^u)^2 m_{\alpha}^2} - \left(\overline{m_{3/2}} y_D^u\right)^2 \nonumber \\
&& \simeq \frac{N M^2}{9} {\tilde m}^2 d_0^2\left(1 + (y_D^u)^2 \right)  \ . \label{scan12}
\end{eqnarray}
The A-terms are such that additional flavor violation (other than from Yukawa couplings) would be from the variance
of the distribution. Thus we have 
\begin{equation}
(\delta_{ij}^u)_{LR/RL} \equiv
{\Delta A^u v_u \over \overline{m_0^2}} \sim   {3 d_0 v_u \over \sqrt{N} \tilde{m} M} \ .  \label{scan14}
\end{equation}
From the above it is clear that, similarly  to the case of scalar masses, there is a suppression  $1/\sqrt{N}$ coming from the large number of hidden-sector fields.

Notice that our starting expressions for soft terms (\ref{soft2}) and the scanning we performed above is different compared to one based on a naive spurion-type parameterization of soft terms:
		
		\begin{equation}
		\label{sugraoperators1}
		\sum_{\alpha=1}^{N}\frac{s_{\alpha }^a}{M_{P}}\int d^2\theta T_{\alpha} W^{a} W^{a}
		\quad  \to \quad \quad
		M^{a}_{1/2} =  \frac{1}{M_{P}}\sum_{\alpha=1}^{N}s_{\alpha }^a F_{T_{\alpha}} \ , 
		\end{equation}
		\begin{equation}
		\label{sugraoperators2}
		\sum_{\alpha=1}^{N}\frac{d_{\alpha, i j}}{M_{P}^2}\int d^4\theta T^{\dagger}_{\alpha}T_{\alpha}Q^{\dagger}_iQ_j 
		\quad \to \quad \quad
		m_{\tilde{f}_{ij}}^2 = \frac{1}{M_{P}^2}\sum_{\alpha=1}^{N}d_{\alpha, ij}F_{T_{\alpha}}^{\dagger}F_{T_{\alpha}} \ , 
		\end{equation}
		\begin{equation}
		\label{sugraoperators3}
		\sum_{\alpha=1}^{N}\frac{a_{\alpha, i jk}}{M_{P}}\int d^2\theta T_{\alpha}Q_iQ_jQ_k
		\quad \to \quad \quad	
		A_{ijk}=\frac{1}{M_{P}}\sum_{\alpha=1}^{N}a_{\alpha, ijk}F_{T_{\alpha}} \ , 
		\end{equation}
		\begin{equation}
		\label{sugraoperators4}
		\sum_{\alpha=1}^{N}\frac{b_{\alpha }}{M_{P}^2}\int d^4\theta T_{\alpha}T^{\dagger}_{\alpha}H_u H_d	
		\quad \to \quad \quad
		B_{H_uH_d}=\frac{1}{M_{P}^2}\sum_{\alpha=1}^{N}b_{\alpha}F_{T_{\alpha}}F_{T_{\alpha}}^{\dagger} \ , 
		\end{equation}
		\begin{equation}
		\label{sugraoperators5}
		\sum_{\alpha=1}^{N}\frac{q_{\alpha }}{M_{P}}\int d^4\theta T^{\dagger}_{\alpha}H_u H_d
		\quad \to \quad \quad
		\mu=\frac{1}{M_{P}}\sum_{\alpha=1}^{N}q_{\alpha}F_{T_{\alpha}}^{\dagger} \ . 	
		\end{equation}
		
		The difference is that in the SUGRA expressions  (\ref{soft2}) the flavor-blind contributions proportional to $m_{3/2}^2$ scan coherently (add up) in soft terms, whereas the other contributions, which
		are similar to the global SUSY expressions (\ref{sugraoperators5}), being multiplied by random couplings scanned around zero, average to zero. A similar scan we performed above, but
		starting from  (\ref{sugraoperators5}) would not lead a suppression of FCNC effects, unlike our scan above. 
						 
	
\section{Numerical Analysis}
Using eqs.(\ref{scan2}) as boundary conditions at the high scale, we perform a numerical analysis of the resulting soft spectrum at the weak scale and studied the phenomenology. 
For the numerical analysis, we have considered $N$ to be 100, with $m_{\alpha}$ varying discretely and randomly from  -100 to 100.  We explored taking $\tilde{m}$ to be 20 GeV.  The maximum value of
$m_{3/2}$ is roughly about 6 TeV.  The parameters $d_\alpha,s_\alpha, a_{\alpha},g_\alpha$ are varied between \{-1/4,1/4\}. A larger
value for the $d_0$ parameters would lead to significant number of the points ruled out due to tachyonic masses
at the weak scale.  We believe that larger values  $d_0 \sim {\cal O}(1)$ would not significantly alter the
results presented here. Finally we set $\tan\beta =10$. We show that these values of $N$ are enough to demonstrate the $1/\sqrt{N}$ suppression on the flavor violating off-diagonal entries. 
We use Suseflav \cite{Chowdhury:2011zr} for computation of the spectrum and computing the flavor
observables.  

\begin{figure}[H]
	\centering
	\subfigure[]{%
		\label{fig:firstb}%
		\includegraphics[width=8cm,height=6cm]{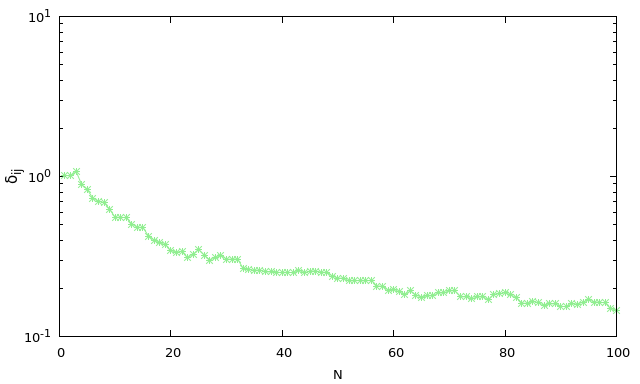}}%
	\hfill
	\subfigure[]{%
		\label{fig:secondb}%
		\includegraphics[width=8cm,height=6cm]{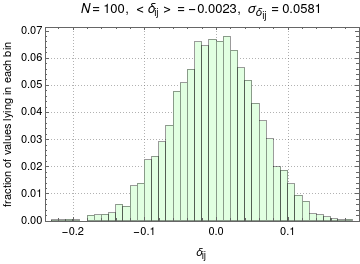}}%
	
	\caption{ A scatter plot showing the variation of maximum value of $\delta_{ij}$  of the type $LL/RR$ 
	with respect to number  of hidden sectors,  $N$ at the high scale (left side). A histogram of the $\delta_{ij}$ is presented.  As expected the mean is very close to zero and the variance is as computed in the text.  }
\end{figure}

The variation of the off-diagonal flavor entries in the sfermion mass matrices is presented in  Figs.(\ref{fig:firstb})
and (\ref{fig:secondb}).  In Fig.(\ref{fig:firstb}) we present the scatter plot of a typical 
$\delta_{ij}$ as defined in eq.(\ref{scan8}).  From the plot it is clear that the $\delta_{ij}$ does fall off as
 $1/\sqrt{N}$. The second figure show the same data in terms of a histogram, where as we can see 
 the mean value is close to zero and the variance is as expected from the formulae in eqs.(\ref{scan7},\ref{scan8}). 

The high scale distributions are then evolved to the weak scale where the full soft supersymmetric 
spectrum is computed. Radiative electroweak symmetry breaking conditions are imposed.  Experimental
constraints from LHC and the Higgs mass are also taken in  consideration. As is standard practice
we consider one $\delta_{ij}$ at a time. In the present letter, we consider the two of the strongest 
constraints, \textit{i.e.} the mass difference between the neutral $K$-mesons, $\Delta M_K$ and 
the leptonic rare decay $\mu \to e+\gamma$. A more detailed analysis with rest of the flavor processes
will be presented elsewhere \cite{edplskv}. 

\begin{figure}[H]
	\centering
	\subfigure[]{%
		\label{deltamkLL}%
		\includegraphics[width=8cm,height=6cm]{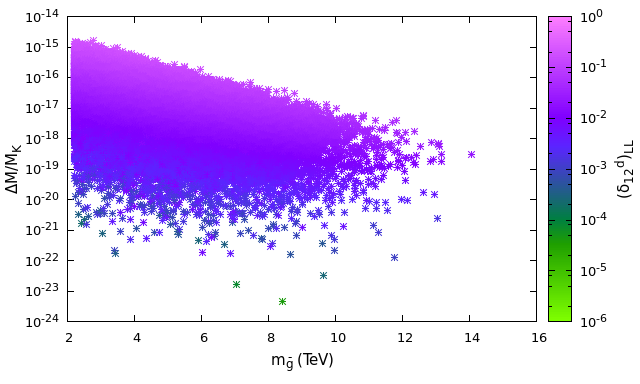}}%
	\hfill
	\subfigure[]{%
		\label{deltamkLR}%
		\includegraphics[width=8cm,height=6cm]{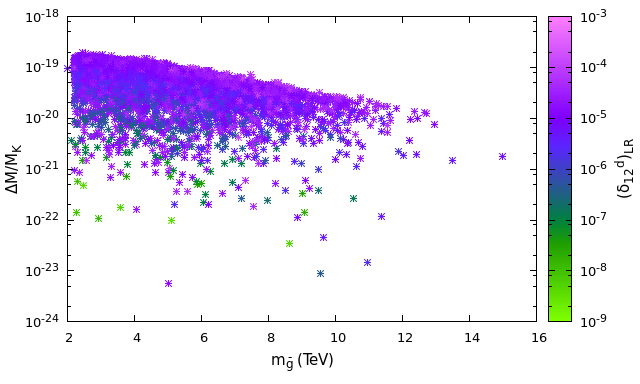}}%
	
	\caption{Regions of the parameter space which satisfy the bounds  from  LHC, Higgs mass and
	other phenomenological bounds. We have chosen $N$ to be 100 and $\tilde{m}$ to be 20 GeV. 
	The distribution of $\delta$ values is as per the 	Eqs. (\ref{scan8}) and (\ref{scan14}).
	  The left side plot is for $LL$ type mass insertion whereas the right hand side is for $LR$ type mass insertion.  All points satisfy the experimental constraint from $\Delta M_K$.}
\end{figure}

At the weak scale, the diagonal entries would be enhanced due to renormalisation group equation
running, while the inter-generational entries of the squark matrices  
 would only receive corrections suppressed by the product of  Yukawa couplings and CKM angles
 \cite{Ciuchini:2007ha}. Due to this the $\delta_{ij}$ would be further suppressed roughly by an additional factor 
 which is proportional to the gluino mass corrections and  roughly independent of the number of 
 hidden sector fields. In figs. \ref{deltamkLL},\ref{deltamkLR} we present the regions of the parameter space allowed by $\Delta M_k$ constraint as a function of the gluino mass. It should be noted here that  we have
 taken the weak scale values of the mass insertions of eq.(\ref{scan8}), 
  where all the parameters appearing on the RHS are computed
 at the weak scale. The  left figure is for the $LL$ mass insertion where as the right figure is for the $LR$ mass
 insertion.  As can be seen from the figure, all the points lie below the experimentally measured value of
 $\Delta M_K$ \cite{Tanabashi:2018oca}.  The spectrum at the weak scale for the first two generations
 is about 5-6 TeV and the gluino mass is shown in the figure after taking into consideration the limits from LHC. 
 For this spectrum and a diluted $\delta \lesssim 10^{-1}$ the constraint from $\Delta M_K$ is satisfied. 
 
 \begin{figure}[H]
	\centering
	\subfigure[]{%
		\label{muegLL}%
		\includegraphics[width=8cm,height=6cm]{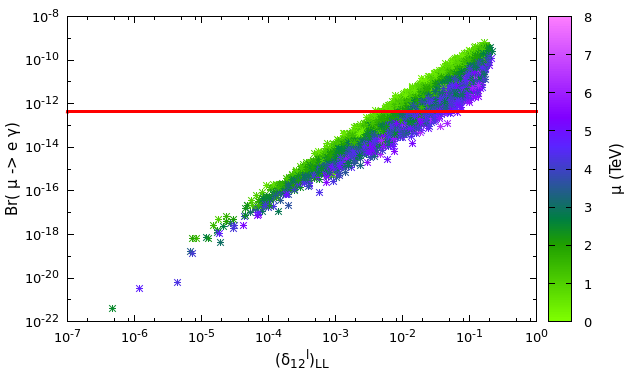}}%
			\hfill
	\subfigure[]{%
		\label{muegLR}%
		\includegraphics[width=8cm,height=6cm]{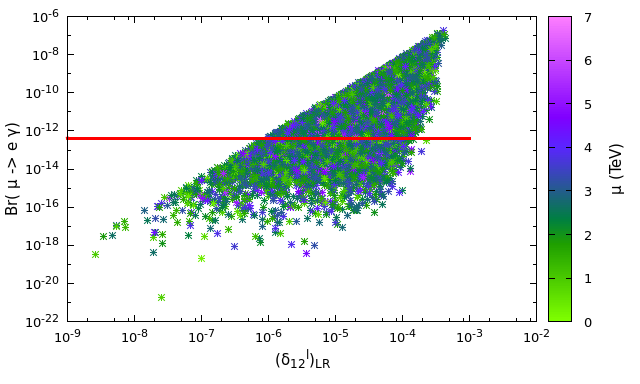}}%
		\caption{ Regions of the parameter space constrained by the leptonic rare decay $\mu \to e + \gamma$
		for $(\delta_{12})_{LL}$ (on the left) and for $(\delta_{12})_{LR}$ (on the right). The horizontal (red) line
		is the present experimental limit from MEG experiment. As can be seen, a large part of the parameter space 
		survives the experimental limit for N=100.}
		\end{figure}
The leptonic rare process  $\mu \to e+\gamma$ is however more strongly constraining for the same set
of parameters, \textit{i.e,} $N =100$ and $\tilde{m} = 20$ GeV. In Figs. \ref{muegLL} and \ref{muegLR}
we present results of the scanning as a function of the $\delta$ parameter and $\mu$. The left figure 
is for a $LL$ type mass insertion whereas the right figure is for $RR$ type mass insertion.  As one can see
from the figures, a significantly large region of the parameter space is still compatible with the latest 
result from the MEG experiment\cite{TheMEG:2016wtm}, but the constraints from  $LR$ are significantly 
stronger, as expected.  A larger $N$ value $\gtrsim 10^5$ would lead to  complete dilution of the $\delta$.

\begin{figure}[H]
	\centering
	\subfigure[]{%
		\label{darkmix}%
		\includegraphics[width=8cm,height=6cm]{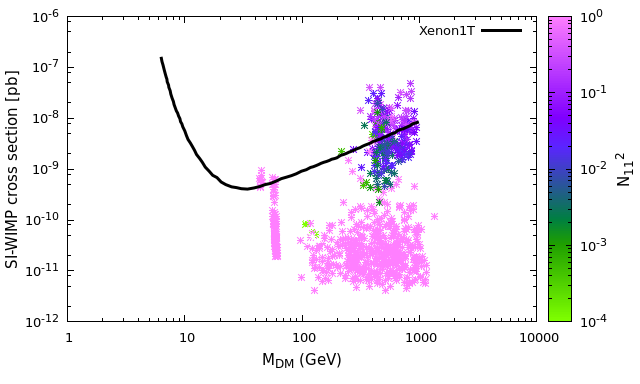}}%
	\hfill
	\subfigure[]{%
		\label{darkcoann}%
		\includegraphics[width=8cm,height=6cm]{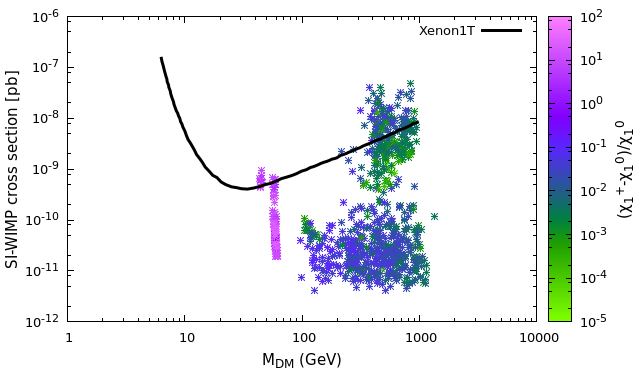}}%
	\caption{Regions of the parameter space which satisfy the relic density and the direct detection results from 
	 Xenon 1T. On the left we have shown the spin independent cross-section with respect to the lightest 
	 neutralino mass and the Bino component of the lightest neutralino. On the right, we show the same, with the
	 mass difference between chargino and Bino. }
\end{figure}

Finally we have also looked for regions with neutralino dark matter which could lead to correct relic density 
while satisfying the constraints from direct detection and flavour.  As can be seen from Fig \ref{darkmix}, 
there  are two branches which satisfy relic density as well as the direct detection result.  The first branch has
 dark matter masses $\lesssim$ 100 GeV and the lightest neutralino is a pure bino. In the second branch 
the neutralino has a region in which it is a pure bino and another region where there is a significant admixture from
wino and  higgsino.  The regions where the neutralino are pure bino have significant co-annihilations with the
chargino as can be seen from the Fig \ref{darkcoann}.  These regions arise due to the non-universality in the gaugino
masses at the high scale due to the $s_\alpha$ parameters.  On the other hand, regions with bino-higgsino mixing
arise due to cancellations in  $M_{1/2}$ in contributions from various fluxes of different spurion fields. 
 As the charges/fluxes ${m}_{\alpha}$ 
take both signs, for significantly large $N$ there is an enhanced probability of 
cancellations between the charges leading to small $M_{1/2}$ at the high scale. We numerically found that 
this probability is significantly high for $N\gtrsim 30$.  Due to the universal 
nature of the gravitino mass, such cancellations do not occur in the 
soft scalar mass terms.  A low value for $\mu$ is very probable in these regions leading to significant 
bino-higgsino mixing. Together they lead to regions with physically viable regions of neutralino dark matter. 
More details of these regions will be presented elsewhere \cite{edplskv}. 

\section{Conclusions and Outlook}

We presented a novel solution to the supersymmetric flavor problem in the presence of large number of 
hidden sector (spurion) fields. Such a scenario naturally arises in the string landscape. The result does not depend on
the explicit details of the string construction, but crucially on the form of the soft terms in the supergravity
potential in the presence of a large number of hidden sector fields, eqs.(\ref{scan2}). They naturally lead to 
a suppression of the flavour violating entries as $1/\sqrt{N}$. At the weak scale, there is further suppression
due to the renormalisation group running, especially for the hadronic mass insertions. 
We have shown that numerically $N=100$ is sufficient to remove the constraints from $\Delta M_K$, whereas  
a much larger $N$ would be required to eliminate completely the constraints on the leptonic sector from
 $\mu \to e + \gamma$. Conversely, a discovery of such leptonic processes in forthcoming experiments could be 
 a smoking gun of such a scenario. 
 The four fluxes contribution to the soft terms presented here provides an interesting framework to further
 study the implications for low energy phenomenology. 

\acknowledgments{
This work is supported by  the CNRS LIA (Laboratoire International Associé) THEP (Theoretical High Energy Physics) and the INFRE-HEPNET (IndoFrench Network on High Energy Physics) of CEFIPRA/IFCPAR (Indo-French Centre for the Promotion of Advanced Research). We also thank  CEFIPRA  for the individual project  grant "Glimpses of New Physics." ED is supported in part by the ANR grant Black-dS-String ANR-16-CE31-0004-01 and PL and SKV are also supported by the Department of Science and Technology, Govt of India  Project "Nature of New Physics". }

	\bibliographystyle{ieeetr}
	\bibliography{NSUSY.bib}	
\end{document}